\documentclass[iop,twocolappendix]{emulateapj}
\usepackage{apjfonts}
\usepackage{amsmath,amssymb}
\usepackage{epstopdf}

\renewcommand{\vec}[1]{\mathbf{#1}}
\newcommand{\diffp}[2]{\frac{\partial #1}{\partial #2}}

\shorttitle{Slow Modes in the Solar Wind}
\shortauthors{Verscharen, Chen, \& Wicks}

\begin{document}


\title{On Kinetic Slow Modes, Fluid Slow Modes, and Pressure-balanced Structures in the Solar Wind} 



\author{Daniel Verscharen$^{1}$, Christopher H.~K.~Chen$^2$, and Robert T.~Wicks$^3$}
\affil{$^{1}$Space Science Center and Department of Physics, University of New Hampshire, Durham, NH 03824, USA; daniel.verscharen@unh.edu\\
$^{2}$Department of Physics, Imperial College London, London SW7 2AZ, UK; christopher.chen@imperial.ac.uk\\
$^{3}$Mullard Space Science Laboratory, University College London, London WC1E 6BT, UK; r.wicks@ucl.ac.uk}

\journalinfo{The Astrophysical Journal, 840:106 (8pp), 2017 May 10}
\submitted{Received 2017 March 6; accepted 2017 March 28; published 2017 May 12}

\begin{abstract}
Observations in the solar wind suggest that the compressive component of inertial-range solar-wind turbulence is dominated by slow modes. The low collisionality of the solar wind allows for nonthermal features to survive, which suggests the requirement of a kinetic plasma description. The least-damped kinetic slow mode is associated with the ion-acoustic (IA) wave and a nonpropagating (NP) mode. We derive analytical expressions for the IA-wave dispersion relation in an anisotropic plasma in the framework of gyrokinetics and then compare them to fully kinetic numerical calculations, results from two-fluid theory, and magnetohydrodynamics (MHD). This comparison shows major discrepancies in the predicted wave phase speeds from MHD and kinetic theory at moderate to high $\beta$. 
MHD and kinetic theory also dictate that all plasma normal modes exhibit a unique signature in terms of their polarization. We quantify the relative amplitude of fluctuations in the three lowest particle velocity moments associated with IA and NP modes in the gyrokinetic limit and compare these predictions with MHD results and in-situ observations of the solar-wind turbulence. 
The agreement between the observations of the wave polarization and our MHD predictions is better than the kinetic predictions, which suggests that the plasma behaves more like a fluid in the solar wind than expected.
\end{abstract}


\section{Introduction}

According to the magnetohydrodynamic (MHD) approximation, a collisional plasma  supports four types of linear modes: the Alfv\'en wave, the fast-magnetosonic wave, the slow-magnetosonic wave, and the entropy mode. While the Alfv\'en wave is noncompressive, the two magnetosonic waves exhibit changes in the density $\delta \rho$ and the magnetic-field strength $\delta |\vec B|$, which are positively correlated in the case of the fast wave and anticorrelated in the case of the slow wave.
According to the more generally valid kinetic theory, a plasma supports an infinite number of linear modes. With a few exceptions, these modes are heavily damped. The Alfv\'en wave and the magnetosonic solutions can still be identified in kinetic theory by their polarizations and dispersion relations \citep{stix92,gary93,klein13}. 

The solar wind is a turbulent, and often collisionless, magnetized plasma that carries a minor component of compressive fluctuations \citep{tu95}. In situ observations indicate that density and field-strength fluctuations in the inertial range are anticorrelated  \citep{bavassano89,yao11,howes12,klein12,yao13,yao13a},  which suggests that the compressions can be usefully modeled as slow modes under the assumption that strong plasma turbulence retains certain characteristics of linear plasma modes \citep{klein12,salem12,chen13,howes14}. Observations also show that the compressive fluctuations are more anisotropic than the Alfv\'enic fluctuations, with $k_{\perp}\gg k_{\parallel}$  \citep{chen12,chen16}, where $k_{\perp}$ ($k_{\parallel}$) is the perpendicular (parallel) component of the wave vector with respect to the background magnetic field.
Some literature interprets part of the compressive fluctuations as pressure-balanced structures (PBSs) \citep[i.e., structures in which the variation in the thermal pressure and the variation in the magnetic pressure balance so that the total pressure stays constant][]{burlaga70,vellante87,burlaga90,zank90,zank93, marsch93,tu94,mccomas95,ghosh98,reisenfeld99, bavassano04,yao11,verscharen12,narita15,yang17}. 
The large collisional mean free paths in the solar wind suggest that a kinetic description of these fluctuations that incorporates effects due to deviations from  thermodynamic equilibrium, such as the commonly observed temperature anisotropies with respect to the background magnetic field,  is necessary \citep{marsch82,kasper02,kasper02a,hellinger06,bale09,chen16a,verscharen16}. 

In kinetic theory, the two modes that are most similar to the slow-magnetosonic mode are the ion-acoustic (IA) wave and a nonpropagating (NP) mode depending on the plasma parameters \citep{howes06}. We refer to these modes as kinetic slow modes. The purpose of this work is to discuss the dispersion relations and the polarization properties of the IA wave, the NP mode, and the MHD slow mode. We use the fluctuations in the three lowest particle velocity moments (density, velocity, and pressure) as observable markers for the polarization of the compressive component of the solar-wind turbulence. By comparing our predictions for these markers with in-situ solar-wind observations, we distinguish between IA-mode-like, NP-mode-like, and MHD-slow-mode-like behavior.
For more details on the IA and NP modes, and a comprehensive derivation of their dispersion relations, we recommend the extensive treatments by \citet{howes06}, \citet{schekochihin09}, and \citet{kunz15}.

\section{Dispersion Relations and Damping Rates of Kinetic Slow Modes}\label{sect_slow_modes}

Assuming large wavelengths ($k_{\perp}\rho_{\mathrm p}\ll 1$), low frequencies ($\omega_{\mathrm r}\ll \Omega_{\mathrm p}$), and $k_{\perp}\gg k_{\parallel}$ in an electron--proton plasma, where $\rho_{\mathrm p}$ is the proton gyroradius, $\omega_{\mathrm r}$ is the real part of the frequency $\omega$, and $\Omega_{\mathrm p}$ is the proton gyrofrequency, the gyrokinetic dispersion relation (see Appendix~\ref{app_slow} for a sketch of the derivation) contains two distinct types of slow modes  \citep{howes06,schekochihin09}. In the limit of low\footnote{We make the assumption that all interspecies temperature ratios, as well as all $R_j$ and $1/R_j$,  are much less than $\sqrt{m_{\mathrm p}/m_{\mathrm e}}$.}  $\beta_{\parallel \mathrm p}$ and low $\beta_{\perp\mathrm p}$, the slow-mode part of the dispersion relation describes IA waves. In the limit of high $\beta_{\parallel\mathrm p}$ and high $\beta_{\perp\mathrm p}$, the slow-mode part of the dispersion relation describes NP modes, where $\beta_{\parallel j}\equiv 8\pi n_{0j}k_{\mathrm B}T_{\parallel j}/B_0^2$, $\beta_{\perp j}\equiv \beta_{\parallel j}R_j$, $n_{0j}$ is the background density of species $j$, and $k_{\mathrm B}$ is the Boltzmann constant. We define the temperature anisotropy of species $j$ as $\Delta_j\equiv R_j-1$, where $R_{j}\equiv T_{\perp j}/T_{\parallel j}$, and $T_{\perp j}$ ($T_{\parallel j}$) is the perpendicular (parallel) temperature of species $j$ with respect to the background magnetic field $\vec B_0$.

\begin{figure}
\includegraphics[width=\columnwidth]{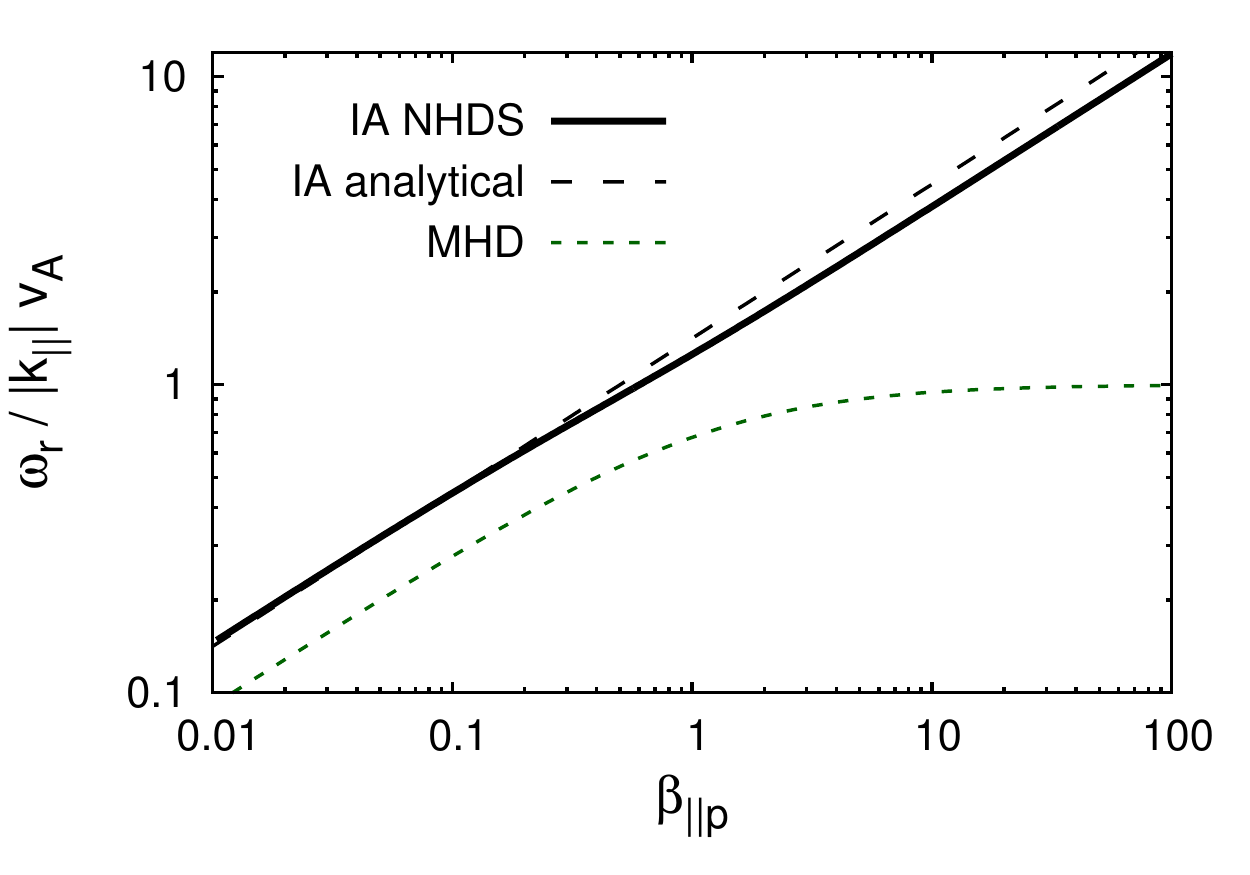}
\caption{Comparison of the phase speeds of the IA wave from the fully kinetic hot-plasma dispersion relation (NHDS) with Equation~(\ref{omegaIA}) in an electron--proton plasma. We use the parameters $T_{\parallel\mathrm p}=T_{\parallel\mathrm e}$, $\theta=88^{\circ}$, $k_{\parallel}v_{\mathrm A}/\Omega_{\mathrm p}=0.001$, and $R_{\mathrm p}=R_{\mathrm e}=1$. The green dashed line shows Equation~(\ref{omegaMHD}) with $\beta=\beta_{\parallel\mathrm p}$ and $\kappa=5/3$. The NP mode fulfills $\omega_{\mathrm r}=0$ exactly.}
\label{fig_dispersion}
\end{figure}

According to our derivation in Appendix~\ref{app_slow}, IA waves fulfill the dispersion relation
\begin{equation}\label{omegaIA}
\omega_{\mathrm r}\simeq k_{\parallel}c_{\mathrm s}, 
\end{equation}
where \citep{stix92,gary93,narita15}
\begin{equation}\label{csIA}
c_{\mathrm s}\equiv \sqrt{\frac{3k_{\mathrm B}T_{\parallel \mathrm p}+k_{\mathrm B}T_{\parallel \mathrm e}}{m_{\mathrm p}}}
\end{equation}
is the IA speed, and $m_{j}$ is the particle mass of species $j$.  As pointed out by \citet{gary93}, a comparison between Equation~(\ref{csIA}) with two-fluid theory, in which $\omega_{\mathrm r}=k_{\parallel}C_{\mathrm F}$ and
\begin{equation}\label{cF}
C_{\mathrm F}\equiv \sqrt{\frac{\kappa_{\mathrm p}k_{\mathrm B}T_{\parallel\mathrm p}+\kappa_{\mathrm e}k_{\mathrm B}T_{\parallel\mathrm e}}{m_{\mathrm p}}},
 \end{equation}
implies that---to the degree to which an adiabatic behavior applies to the kinetic solution---the specific heat ratios in IA waves fulfill $\kappa_{\mathrm p}=3$ and $\kappa_{\mathrm e}=1$ for protons and electrons, respectively. From this point of view, protons behave like a one-dimensional adiabatic component due to their degree of freedom along $\vec B_0$, while electrons behave like an isothermal component due to their large thermal speed compared to the wave phase speed.
The imaginary part $\gamma$ of the IA-wave frequency is given by
\begin{equation}\label{gammaIA}
\gamma\simeq-|k_{\parallel}|c_{\mathrm s}\sqrt{\pi}\frac{c_{\mathrm s}^3}{w_{\parallel\mathrm p}^3}\frac{e^{-c_{\mathrm s}^2/w_{\parallel\mathrm p}^2}}{1+3w_{\parallel\mathrm p}^2/c_{\mathrm s}^2},
\end{equation}
where $w_{\parallel\mathrm p}\equiv \sqrt{2k_{\mathrm B}T_{\parallel\mathrm p}/m_{\mathrm p}}$ is the parallel thermal speed of the protons. The derivation of Equations~(\ref{omegaIA}) and (\ref{gammaIA}) requires the assumption that $T_{\parallel\mathrm e}\gg T_{\parallel\mathrm p}$ so that $\gamma\ll\omega_{\mathrm r}$, although a comparison with numerical results in Figure~\ref{fig_dispersion} will show that Equation~(\ref{omegaIA}) is a good approximation  for even $T_{\parallel\mathrm e}\approx T_{\parallel\mathrm p}$.
Equation~(\ref{gammaIA}) reduces to the result given by \citet{howes06} for an isotropic plasma with $w_{\parallel\mathrm p}\ll c_{\mathrm s}$. The approximated IA dispersion relation in Equations~(\ref{omegaIA}) and (\ref{gammaIA}) does not depend on $T_{\perp j}$ since the restoring force in the IA wave is solely due to the parallel pressure gradients of the protons and electrons \citep[also see][]{basu08}. 

According to our derivation in Appendix~\ref{app_slow}, NP modes fulfill the dispersion relation \citep[also see][]{foote79,howes06,kunz15} $\omega_{\mathrm r}=0$ and  
\begin{equation}\label{gammaNP}
\gamma\simeq -\frac{|k_{\parallel}|v_{\mathrm A}}{R_{\mathrm p}^2\sqrt{\pi \beta_{\parallel\mathrm p}}}\left(1-\beta_{\perp\mathrm p}\Delta_{\mathrm p}-\beta_{\perp\mathrm e}\Delta_{\mathrm e}\right),
\end{equation}
where $v_{\mathrm A}\equiv B_0/\sqrt{4\pi n_{0\mathrm p}m_{\mathrm p}}$ is the proton Alfv\'en speed. The NP mode can become unstable ($\gamma>0$) according to Equation~(\ref{gammaNP}) if 
\begin{equation}\label{mirror}
\beta_{\perp\mathrm p}\Delta_{\mathrm p}+\beta_{\perp\mathrm e}\Delta_{\mathrm e}>1,
\end{equation}
 which is the mirror-mode instability criterion.

For the sake of brevity, we limit ourselves to the isotropic case for the following numerical evaluations of the dispersion relation.
In Figure~\ref{fig_dispersion}, we compare Equation~(\ref{omegaIA}) with numerical results of the fully kinetic hot-plasma dispersion relation obtained with the numerical code NHDS \citep{verscharen13a}. For reference, we show the MHD slow-mode dispersion relation,
\begin{equation}\label{omegaMHD}
\omega=k_{\parallel}v_{\mathrm A}\frac{C_-}{\cos\theta},
\end{equation}
where
\begin{equation}\label{Cminus}
C_{-}\equiv \sqrt{\frac{1}{2}\left(1+\frac{\kappa}{2}\beta\right)- \frac{1}{2}\left[\left(1+\frac{\kappa}{2}\beta\right)^2-2\kappa \beta\cos^2\theta\right]^{1/2}}
\end{equation}
is the slow-magnetosonic speed, $\beta$ is the ratio of thermal to magnetic pressure, and $\kappa$ is the specific heat ratio. We set $\beta=\beta_{\parallel\mathrm p}$ and $\kappa=5/3$. We denote the angle between $\vec k$ and $\vec B_0$ as $\theta$. 
While the numerical and the analytical dispersion relations for the IA wave agree well with each other (even at high $\beta_{\parallel\mathrm p}$), the MHD solution shows a significant deviation from the kinetic solutions, especially at high $\beta_{\parallel\mathrm p}$.

\begin{figure}
\includegraphics[width=\columnwidth]{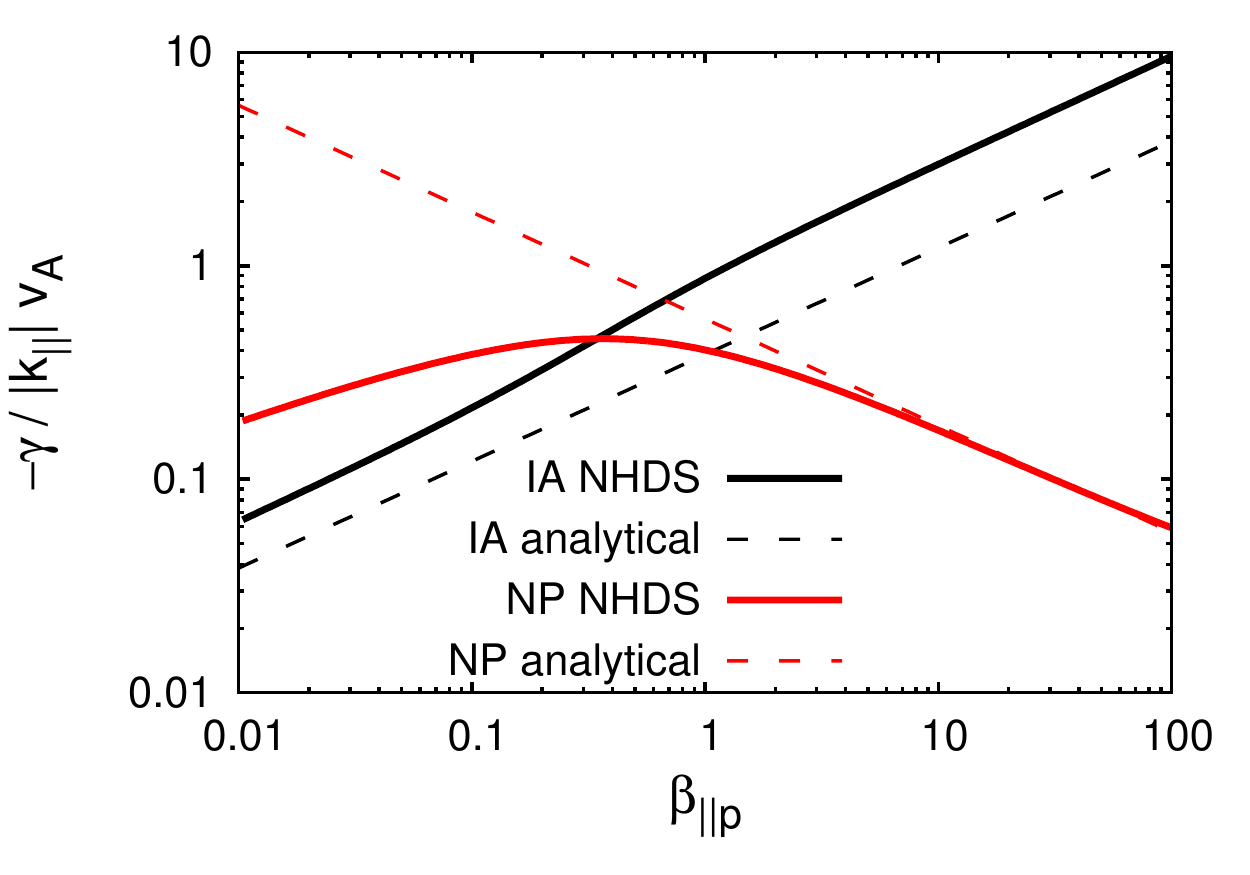}
\caption{Damping rates of the IA wave and the NP mode in numerical solutions obtained with NHDS and according to Equations~(\ref{gammaIA}) and (\ref{gammaNP}). The parameters are the same as those in Figure~\ref{fig_dispersion}.}
\label{fig_damping_rates}
\end{figure}

In Figure~\ref{fig_damping_rates}, we compare  Equations~(\ref{gammaIA}) and (\ref{gammaNP}) with NHDS solutions. For this parameter set, the analytical results and the numerical results agree well for the NP mode at high $\beta_{\parallel \mathrm p}$, as assumed in the derivation of Equation~(\ref{gammaNP}). Even at low $\beta_{\parallel\mathrm p}$,  Equation~(\ref{gammaIA}) and the numerical IA solution deviate from each other, which is attributed to finite-$|\gamma/\omega_{\mathrm r}|$ effects. Using this parameter set, the numerical damping rates of both modes are equal at $\beta_{\parallel \mathrm p}\approx 0.3$. 

\section{Polarization Properties and Relation to Pressure-balanced Structures}\label{polarization}

We use the distinct polarization properties of the IA, NP, and MHD slow modes to identify the dominating mode in the solar wind. For that purpose, we define the dimensionless quantities $\xi_j$, $\chi_j$, $\alpha_{\perp j}$, and $\alpha_{\parallel j}$ as the normalized observable amplitudes of fluctuations in the three lowest velocity moments:
\begin{equation}\label{xi}
\frac{\delta n_j}{n_{0j}}=\xi_j \frac{\delta B_{\parallel}}{B_0},
\end{equation}
where $\delta n_j$ is the amplitude of fluctuations in the particle number density;
\begin{equation}\label{chi}
\frac{\delta U_{\parallel j}}{v_{\mathrm A}}=\chi_j \frac{\delta B_{\parallel}}{B_0},
\end{equation}
where $\delta U_{\parallel j}$ is the amplitude of fluctuations in the $\vec B_0$-parallel bulk velocity; and
\begin{align}
\frac{\delta p_{\perp j}}{p_{B0}}&=\alpha_{\perp j} \frac{\delta B_{\parallel}}{B_0},\\
\frac{\delta p_{\parallel j}}{p_{B0}}&=\alpha_{\parallel j} \frac{\delta B_{\parallel}}{B_0},
\end{align}
where $\delta p_{\perp j}$ ($\delta p_{\parallel j}$) is the amplitude of fluctuations in the thermal pressure in the direction that is perpendicular (parallel) to $\vec B_0$, and $p_{B0}\equiv B_0^2/8\pi$.
We derive expressions for  $\xi_j$, $\chi_j$, $\alpha_{\perp j}$, and $\alpha_{\parallel j}$ in the gyrokinetic limit in Appendix \ref{app_PBS} and give the MHD equivalents in Appendix~\ref{app_MHD}. 
Although all of these fluctuating moments are observable identifiers for the underlying plasma modes, the fluctuations in the total pressure play a decisive role due to the prominent observations of PBSs in the solar wind. Therefore, we define the normalized amplitude $\psi_j$ of fluctuations in the thermal pressure  as
\begin{equation}\label{psi}
\frac{\delta p_j}{p_{B0}}=\psi_j \frac{\delta B_{\parallel}}{B_0}.
\end{equation}
With $p_{j}\equiv (2p_{\perp j}+p_{\parallel j})/3$, we find
\begin{equation}\label{psieq}
\psi_j= \frac{2}{3}\alpha_{\perp j}+\frac{1}{3}\alpha_{\parallel j}.
\end{equation}

In MHD, PBSs are associated with the MHD slow mode in the quasi-perpendicular limit \citep{tu95,kellogg05,klein13}, which suggests that the IA wave and the NP mode are associated with the kinetic counterparts of PBSs. 
All wave solutions in the gyrokinetic approximation fulfill the perpendicular pressure balance according to Equation~(\ref{gyroPB}). However, the pressure balance commonly associated with PBSs implies that the sum of the total thermal pressures and the magnetic pressure remain constant:
\begin{equation}\label{PB}
p_{\mathrm p}+p_{\mathrm e}+p_B=\mathrm{constant},
\end{equation}
where  $p_B\equiv |\vec B|^2/8\pi$. 
Linearizing Equation~(\ref{PB}) leads to
\begin{equation}
\frac{\delta p_{\mathrm p}}{p_{B0}}+\frac{\delta p_{\mathrm e}}{p_{B0}}+\frac{\delta p_B}{B_0}=0.
\end{equation}
With $\delta p_B\approx B_0 \delta B_{\parallel}/4\pi$,  pressure balance according to Equation~(\ref{PB}) is then achieved if
\begin{equation}\label{APB}
\psi\equiv \psi_{\mathrm p}+\psi_{\mathrm e}=-2.
\end{equation}

The quantities $\xi_j$, $\chi_j$, $\alpha_{\perp j}$, $\alpha_{\parallel j}$, $\psi_j$, and $\psi$ are complex quantities. For their comparison with observations in Section~\ref{sect_results}, we express them in terms of their magnitude $|\cdot|$  and phase angle $\mathrm{arg}(\cdot)$. 

\subsection{Observational Methods}

For the observational analysis, we use data from the \emph{WIND} spacecraft \citep{acuna95} in the solar wind at 1 au. We use magnetic-field data from the MFI instrument \citep{lepping95} at 3 s resolution. For the particles, we use the ground-calculated moments from the 3DP instrument \citep{lin95}, at 24 s resolution for the ions and at 98 s resolution for the electrons. We split the data from the period 2004 July 1 to 2014 December 31 into non-overlapping 1 hr intervals and then linearly interpolate data gaps. We exclude intervals with data gaps greater than 5\% from the analysis. 

In each 1 hr interval, we determine the mean and fluctuating values in order to calculate the polarization properties $\xi_{\mathrm p}$, $\chi_{\mathrm p}$, and $\psi$, as defined in Section~\ref{polarization}. We determine $n_{0\mathrm p}$ and $\vec B_0$  as averages over each interval and determine $v_{\mathrm A}$ and $p_{B0}$ from those averages. We also determine $\beta_{\parallel\mathrm p}$ from the average density, magnetic field, and parallel temperature of the interval. For the amplitudes of $\xi_{\mathrm p}$, $\chi_{\mathrm p}$, and $\psi$, we calculate the rms values of each quantity. For the phases of $\xi_{\mathrm p}$, $\chi_{\mathrm p}$, and $\psi$, we calculate the wavelet coherence spectrum  \citep{torrence98},
\begin{equation}
C_{\delta A,\delta | \vec B|/B_0}(a,b)\equiv S\left[W_{\delta A}^{\ast}(a,b)\,\, W_{\delta |\vec B|/B_0}(a,b)\right],
\end{equation}
 for each quantity $\delta A \in (\delta n_{\mathrm p}/n_{0\mathrm p}, \delta U_{\parallel \mathrm p}/v_{\mathrm A}, \delta p_{\perp j}/p_{B0},\delta p_{\parallel j}/p_{B0})$, where $W_X(a,b)$ is the continuous Morlet-wavelet transform of $X$ at scales $a$ and positions $b$, and $S$ is a smoothing operator in time and scale. The phase is defined as the coherence phase,
 \begin{equation}
 \text{phase}=\tan^{-1}\frac{\mathrm{Im}\left[C_{\delta A,\delta| \vec B|/B_0}(a,b)\right]}{\mathrm{Re}\left[C_{\delta A,\delta| \vec B|/B_0}(a,b)\right]}.
 \end{equation}
We then take the average over scales between 1 hr and 15 minutes. At shorter timescales, instrument noise would contaminate our results. In our data analysis, we use $\delta |\vec B|$ instead of $\delta B_{\parallel}$ in the definitions of $\xi_{\mathrm p}$, $\chi_{\mathrm p}$, and $\psi$ in order to reduce uncertainties in the determination of the $\vec B_0$-parallel direction. In the limit in which our linear analysis is valid, $\delta |\vec B|\simeq \delta B_{\parallel}$. We then create two-dimensional histograms in the $\xi_{\mathrm p}$-$\beta_{\parallel\mathrm p}$, $\chi_{\mathrm p}$-$\beta_{\parallel\mathrm p}$, and $\psi$-$\beta_{\parallel\mathrm p}$ planes, and then bin the data logarithmically in $\beta_{\parallel \mathrm p}$ and $\xi_{\mathrm p}$, and linearly in $\chi_{\mathrm p}$ and $\psi$. Last, we normalize each column of data in the histogram at a constant $\beta_{\parallel\mathrm p}$ by the peak counts in that column so that each column is independently normalized to show the peak in the variable at a fixed $\beta_{\parallel\mathrm p}$.

\subsection{Results}\label{sect_results}

According to observations \citep[e.g., ][]{kasper02,kasper02a,hellinger06,bale09,chen16a}, the mirror-mode and firehose thresholds set approximate upper and lower limits on $R_{\mathrm p}$. Therefore, we set the maximum and minimum values of $R_{\mathrm p}$ in our calculations to the values that fulfill Equation~(\ref{mirror}) (labeled as ``IA/NP mirror'') or Equation~(\ref{firehose}) (labeled as ``NP firehose''), after replacing the inequality signs with equality signs. These curves identify the extreme values for $\xi_{\mathrm p}$, $\chi_{\mathrm p}$, and $\psi$ in an anisotropic plasma. For a comparison, we also include the results from isotropic MHD as derived in Appendix~\ref{app_MHD}.

\begin{figure}
\includegraphics[width=\columnwidth]{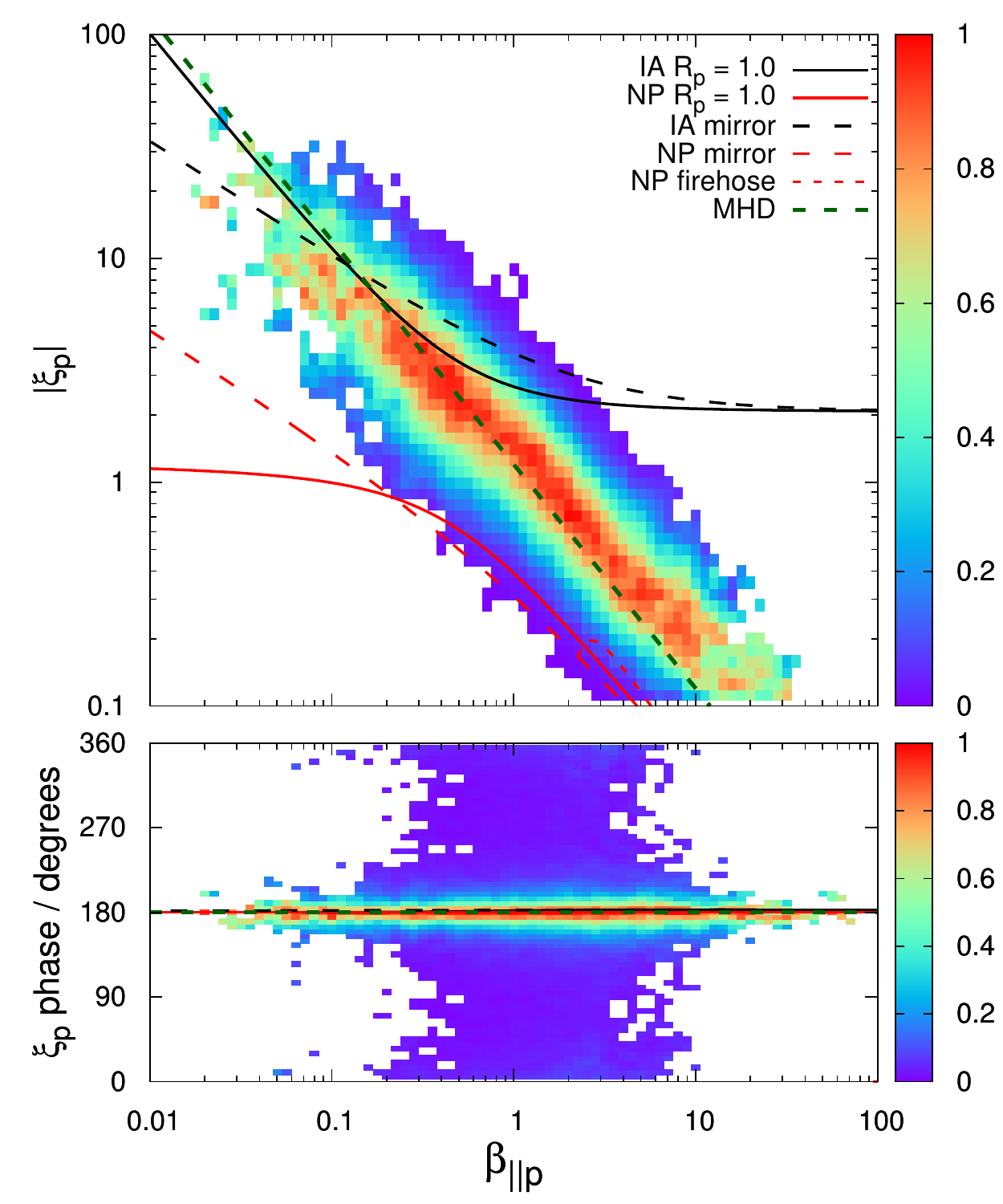}
\caption{$\xi_{\mathrm p}$ as a function of $\beta_{\parallel\mathrm p}$. The lines show our theoretical results and the color-coded dots show the linearly-scaled column-normalized data distribution in the $\xi_{\mathrm p}$-$\beta_{\parallel\mathrm p}$ plane. The top panel shows $|\xi_{\mathrm p}|$, and the bottom panel shows the phase angle between $\delta n_{\mathrm p}$ and $\delta B_{\parallel}$.}
\label{fig_xi}
\end{figure}

We show the numerical and observational results for the zeroth velocity moment ($\xi_{\mathrm p}$) as functions of $\beta_{\parallel\mathrm p}$ in Figure~\ref{fig_xi}. 
Our theoretical results indicate that the vast majority of the data exhibit phase angles of $\sim 180^{\circ}$, which shows a strong anticorrelation between $\delta n_{\mathrm p}$ and $\delta B_{\parallel}$. The MHD prediction is shown for $\kappa=5/3$. Although the NP-mode prediction agrees better with the observations at large $\beta_{\parallel\mathrm p}$ than the IA-wave prediction and vice versa, the MHD solution shows the best agreement overall with the observations of $\xi_{\mathrm p}$ in both the magnitude and phase. In the highly oblique limit, Equation~(\ref{xiMHD}) leads to $\xi_{\mathrm{MHD}}\rightarrow -2/(\kappa \beta_{\parallel\mathrm p})$. A best fit to the data in Figure~\ref{fig_xi} using this limit shows that the observed $\xi_{\mathrm p}$-behavior of the large-scale compressive fluctuations in the solar wind corresponds to the MHD behavior of highly oblique slow modes with
\begin{equation}
\kappa=1.4412\pm0.0036,
\end{equation}
where the error margin represents the statistical error of the fit only.

\begin{figure}
\includegraphics[width=\columnwidth]{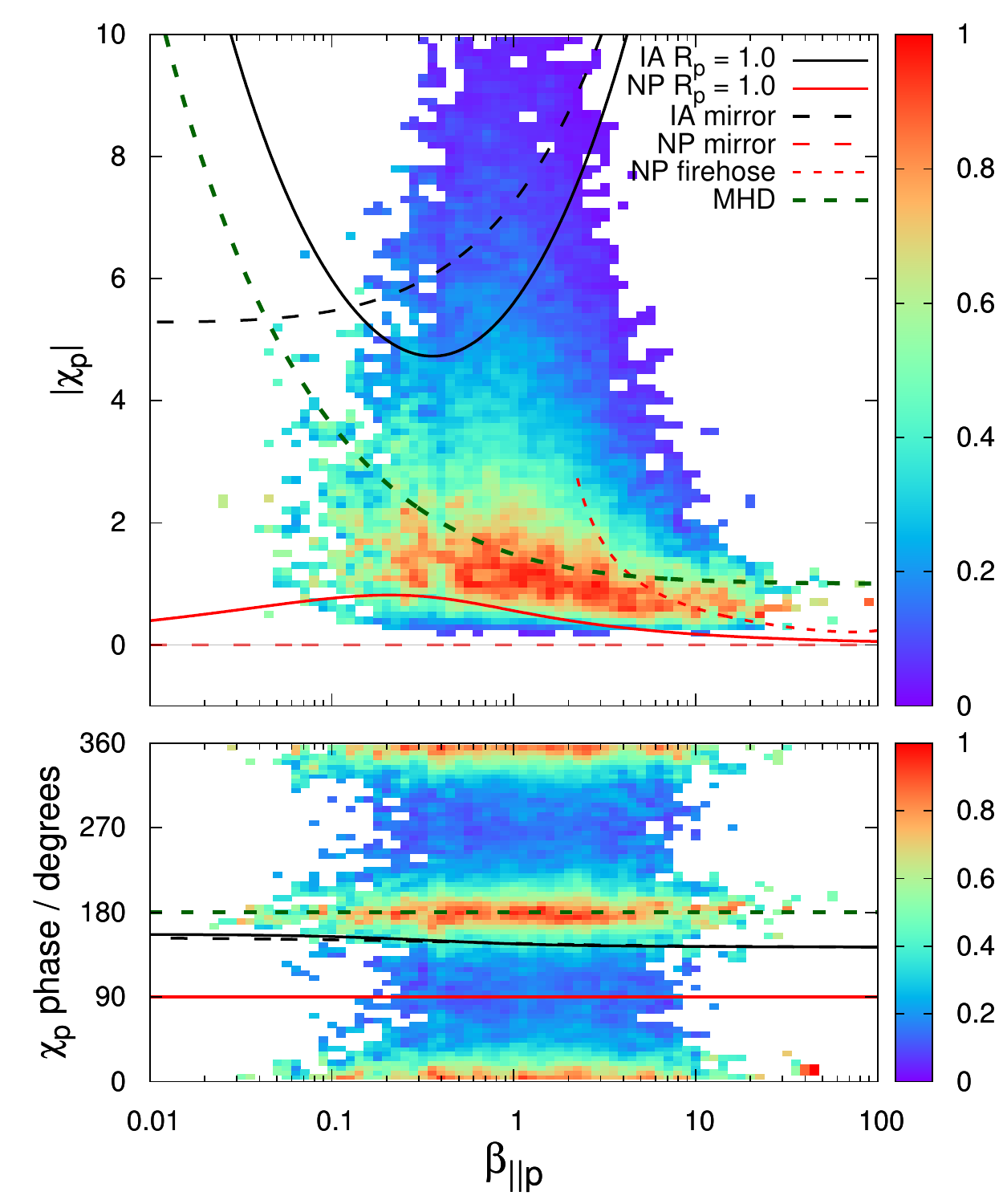}
\caption{$\chi_{\mathrm p}$ as a function of $\beta_{\parallel\mathrm p}$. The lines show our theoretical results and the color-coded dots show the linearly-scaled column-normalized data distribution in the $\chi_{\mathrm p}$-$\beta_{\parallel\mathrm p}$ plane. The top panel shows $|\chi_{\mathrm p}|$, and the bottom panel shows the phase angle between $\delta U_{\parallel \mathrm p}$ and $\delta B_{\parallel}$.}
\label{fig_chi}
\end{figure}

We compare our predictions for the first velocity moment ($\chi_{\mathrm p}$) with observations in Figure~\ref{fig_chi}.
Our theoretical results show that $|\chi_{\mathrm p}|$ is much greater in the IA wave than in the NP mode. While the theoretical NP-mode solutions for $R_{\mathrm p}=1$ and at the firehose threshold exhibit a $90^{\circ}$-phase shift between $\delta U_{\parallel\mathrm p}$ and $\delta B_{\parallel}$, the observations predominantly exhibit a phase shift of $\sim 180^{\circ}$ and $\sim 0^{\circ}$. As in the case of $\xi_{\mathrm p}$, the MHD solution shows the best agreement with the observations of $\chi_{\mathrm p}$ in both the magnitude and phase. We note, however, that the measurement of $\chi_{\mathrm p}$ is prone to Alfv\'enic leakage, which increases the uncertainty of this observation. Alfv\'enic leakage is a result of the dominant Alfv\'enic fluctuations and their characteristic (anti-)correlation between $\delta \vec B$ and $\delta \vec U_{\mathrm p}$. Fluctuations on time scales comparable to the time scale we use in defining $\vec B_0$ introduce inaccuracies to our projections of $\delta \vec B$ and $\delta \vec U_{\mathrm p}$ onto the $\vec B_0$-parallel direction. Consequently, some of the strong transversal Alfv\'enic fluctuations in $\delta \vec B$ and $\delta \vec U_{\mathrm p}$ appear as partly field-parallel. Alfv\'enic leakage creates signals at phases of both $\sim 180^{\circ}$ and $\sim 0^{\circ}$.

\begin{figure}
\includegraphics[width=\columnwidth]{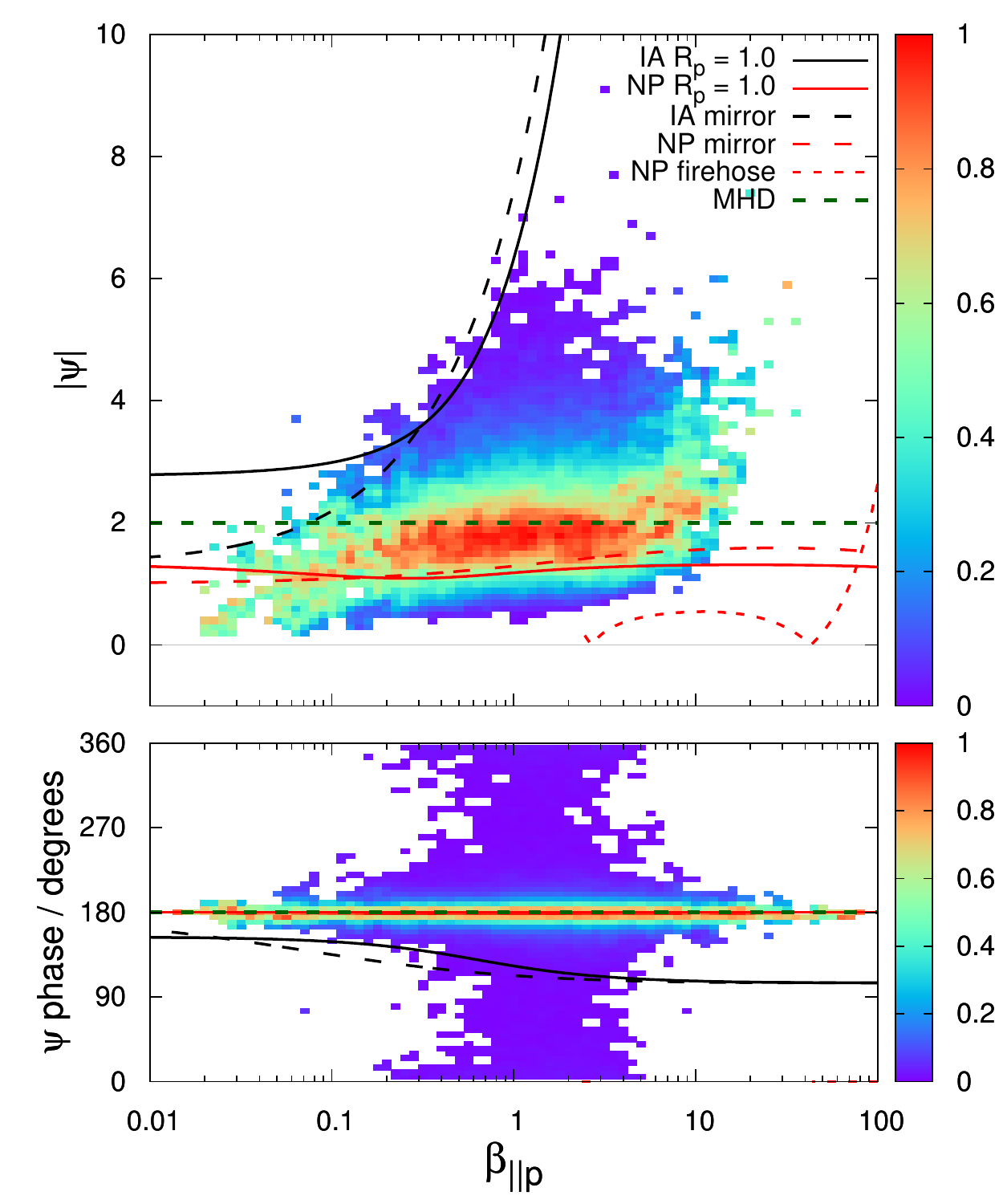}
\caption{$\psi$ as a function of $\beta_{\parallel\mathrm p}$. The lines show our theoretical results and the color-coded dots show the linearly-scaled column-normalized data distribution in the $\psi$-$\beta_{\parallel\mathrm p}$ plane. The top panel shows $|\psi|$, and the bottom panel shows the phase angle between $(\delta p_{\mathrm p}+\delta p_{\mathrm e})$ and $\delta B_{\parallel}$.}
\label{fig_psi}
\end{figure}

Lastly, we show our results for $\psi$ in Figure~\ref{fig_psi}. 
The IA wave and the NP mode do not fulfill Equation~(\ref{APB}). 
The MHD prediction exhibits a full pressure balance and a reasonable agreement with the observations, especially at $\beta_{\parallel\mathrm p}\gtrsim 0.5$.


\section{Discussion and Conclusions}

The long-wavelength kinetic slow mode is associated with two types of compressive modes: the IA wave and the NP mode, both of which exhibit an anticorrelation between $\delta n_j$ and $\delta |\vec B|$. A comparison of the damping rates of the IA mode and the NP mode suggests that the IA mode is the dominating kinetic slow mode at low $\beta_{\parallel\mathrm p}$, while the NP mode is the dominating kinetic slow mode at high $\beta_{\parallel \mathrm p}$.
Temperature anisotropies alter the dispersion relations and the damping behavior of slow modes in kinetic plasmas, and can drive the NP mode (i.e., the mirror mode) to be unstable. 

While MHD does not account for the NP mode\footnote{We note that the MHD entropy mode shares certain characteristics with the NP kinetic slow mode; however, there are significant differences between these modes. For example, the MHD entropy mode does not have fluctuations in $\vec B$.} or the kinetic damping of slow modes, the $\beta_{\parallel\mathrm p}$-dependence of $\omega_{\mathrm r}$ for the IA wave is roughly represented by the MHD dispersion relation at low $\beta_{\parallel\mathrm p}$. At high $\beta_{\parallel\mathrm p}$, however, the MHD approximation does not agree with the dispersion relations of either type of kinetic slow modes. 
Our careful comparison of theoretical predictions for the three lowest particle velocity moments associated with IA, NP, and MHD slow modes with observations suggests that the compressive component of the solar-wind fluctuations is not IA-wave- or NP-mode-like. Our MHD slow-mode  calculation predicts the observed $\beta_{\parallel\mathrm p}$-dependence of $\psi$ well, which corroborates the notion that the MHD slow mode is a potential candidate for PBSs observed in the solar wind \citep{kellogg05,yao13,yao13a}. 

\citet{klein12} compare the normalized correlation between $\delta n_{\mathrm p}$ and $\delta B_{\parallel}$ in a synthetic spacecraft dataset based on a combination of slow and fast modes with observations. In their analysis, the synthetic data based on a superposition of critically-balanced IA waves explain the observed $\beta_{\parallel\mathrm p}$-dependence of the normalized correlation for $0.1\leq \beta_{\parallel\mathrm p}\leq 10$ better than the synthetic data based on a superposition of MHD waves. Applying  \citeauthor{klein12}'s \citeyearpar{klein12} synthetic spacecraft data method  to our markers $\xi_{\mathrm p}$, $\chi_{\mathrm p}$, and $\psi$ over a wider $\beta_{\parallel\mathrm p}$-range using a composition of IA, NP, and other kinetic and MHD modes may lead to observable predictions that clarify the nature of the compressive component of the solar-wind turbulence. In addition, appropriate fully nonlinear turbulence simulations can determine the behavior of slow modes in a turbulent background. However, these studies are beyond the scope of our work.

Although the large collisional mean free paths in the solar wind suggest a preference of kinetic models over MHD models, our observations agree better with the MHD solutions than with the IA-wave or NP-mode solutions.  Our study, therefore, suggests that the large-scale compressive fluctuations in the solar wind behave more fluid-like than kinetic-slow-mode-like. Even the restriction of our data to measurements with low collisional age, $A_{\mathrm c}<0.1$ \citep{kasper08,bourouaine11}, or a reduction of the average time from 1 hr intervals to 10 minute intervals does not change this result (plots for these restrictions are not shown). This discovery suggests that some fundamental process, which  remains to be identified, creates an effective collisionality and thus the requirements for the application of the MHD framework. Possible candidates for such processes include the inhibition of the development of fine structures and damping due to anti-phasemixing \citep[see Section 5.3 in][]{schekochihin16}, scattering by wave--particle collisions and kinetic instabilities  \citep{riquelme15,kunz16,riquelme16,verscharen16,yoon16}, or the isotropization due to the dissipation of electric fluctuations \citep{bale05}. These processes may be scale-dependent, so that a faster measurement cadence could reveal a more kinetic-slow-mode-like behavior on short time scales. The instrumentation on the upcoming missions Solar Orbiter and Solar Probe Plus will provide appropriate measurements for such a scale-dependent study.

\acknowledgments
We thank Ben Chandran and Kris Klein for very helpful discussions. D.V. is supported by NSF/SHINE grant AGS-1460190 and NASA grant NNX16AG81G. C.H.K.C. is supported by an STFC Ernest Rutherford Fellowship.

\appendix
\section{Appendix A\\Derivation of the Dispersion Relation in the Gyrokinetic Approximation}\label{app_slow}

In this appendix, we sketch the derivation of the gyrokinetic slow-mode dispersion relation in an anisotropic plasma combining the derivations given by \citet{howes06} and \citet{kunz15} and using their notation. For simplicity, we assume a bi-Maxwellian background distribution function:
\begin{equation}
f_{0j}=\frac{n_{0j}}{\pi^{3/2}w_{\perp j}^2w_{\parallel j}}\exp\left(-\frac{v_{\perp}^2}{w_{\perp j}^2}-\frac{v_{\parallel}^2}{w_{\parallel j}^2}\right),
\end{equation}
where $w_{\perp j}\equiv\sqrt{2k_{\mathrm B}T_{\perp j}/m_j}$. In the gyrokinetic ordering ($\epsilon\sim k_{\parallel}/k_{\perp}\sim \omega/\Omega_{\mathrm p}\sim\dots \ll 1$), the distribution function is decomposed as 
\begin{equation}
f_j=f_{0j}+\delta f_{1j}+h_j+\dots,
\end{equation}
where
\begin{equation}
\delta f_{1j}\equiv -\frac{q_j}{k_{\mathrm B}T_{\perp j}}\left(\phi+\frac{v_{\parallel}A_{\parallel}}{c}\Delta _j\right)f_{0j}
\end{equation}
is the Boltzmann response, $h_j$ is the gyrokinetic response, $q_j$ is the charge of species $j$, $\phi$ is the electrostatic potential, $\vec A$ is the vector potential, and $c$ is the speed of light.
The collisionless gyrokinetic equation to first order in $\epsilon$ is then given by
\begin{multline}
\diffp{h_j}{t}+v_{\parallel}\diffp{h_j}{z}+\frac{c}{B_0}\left\{\left<\chi \right>_{\vec R},h_j\right\}\\
=\frac{q_jf_{0j}}{k_{\mathrm B}T_{\perp j}}\diffp{\left<\chi\right>_{\vec R}}{t}-\frac{q_jf_{0j}\Delta_j}{k_{\mathrm B}T_{\perp j}}v_{\parallel}\diffp{\left<\chi\right>_{\vec R}}{z},
\end{multline}
where 
$\left<\chi\right>_{\vec R}\equiv \left<\phi-\vec v\cdot \vec A/c\right>_{\vec R}$, and $\left\{\cdot,\cdot\right\}$ is the Poisson bracket. The symbol $\left<\cdot \right>_{\vec R}$ indicates the ring average at the fixed gyrocenter $\vec R$ of species $j$.

Applying a plane-wave ansatz to $h_j$ allows us to express the ring average in terms of Bessel functions of order $m$, $J_{mj}\equiv J_m(k_{\perp}v_{\perp}/\Omega_{j})$, which yields 
\begin{multline}
h_j=\frac{q_jf_{0j}}{k_{\mathrm B}T_{\perp j}}\left\{\left(\omega+k_{\parallel}v_{\parallel}\Delta_j\right) J_{0j}\frac{A_{\parallel}}{k_{\parallel}c}+\left(\frac{\omega R_j}{\omega-k_{\parallel}v_{\parallel}}-\Delta_j \right)\right.\\
\left.\times \left[J_{0j}\left(\phi-\frac{\omega A_{\parallel}}{k_{\parallel} c}\right)+\frac{k_{\mathrm B}T_{\perp j}}{q_j}\frac{2v_{\perp}^2}{w_{\perp j}^2}\frac{J_{1j}\Omega_j}{k_{\perp}v_{\perp}}\frac{\delta B_{\parallel}}{B_0}\right]\right\}.
\end{multline}
The three lowest velocity moments of the fluctuating gyrokinetic distribution function to first order (i.e., the Boltzmann response and the gyrokinetic response) describe the fluctuations in density, bulk velocity, and the perpendicular and parallel thermal pressures as:
\begin{multline}
\delta n_{j}=\frac{q_{j}n_{0j}}{k_{\mathrm B}T_{\perp j}}\left\{\vphantom{\frac{A}{A}}\Gamma_{0j}\frac{\omega A_{\parallel}}{k_{\parallel}c}-\phi-\left(R_{j}\zeta_{j} Z_{j}+\Delta_{j}\right)\right.\\
\times \left.\left[\Gamma_{0j}\left(\phi-\frac{\omega A_{\parallel}}{k_{\parallel}c}\right)+\Gamma_{1j}\frac{k_{\mathrm B}T_{\perp  j}}{q_{j}}\frac{\delta B_{\parallel}}{B_0}\right]\right\}, \label{nmoment} 
\end{multline}
\begin{multline}
\delta U_{\parallel j}=\frac{q_j}{k_{\mathrm B}T_{\perp j}}\left\{\vphantom{\frac{A}{A}}\frac{w_{\parallel j}^2A_{\parallel}}{2c}\Delta_j \left(\Gamma_{0j}-1\right)-\frac{\omega}{k_{\parallel}}R_j\left(\zeta_j Z_j+1\right)\right.\\
\times\left.\left[\Gamma_{0j}\left(\phi-\frac{\omega A_{\parallel}}{k_{\parallel}c}\right)+\Gamma_{1j}\frac{k_{\mathrm B}T_{\perp j}}{q_j}\frac{\delta B_{\parallel}}{B_0}\right]\right\}, \label{Umoment} 
\end{multline}
\begin{multline}
\delta p_{\perp j}=q_{j}n_{0j}\left\{\vphantom{\frac{A}{A}}\Gamma_{3j}\frac{\omega A_{\parallel}}{k_{\parallel}c}-\phi-\left(R_{j }\zeta_{j } Z_{j}+\Delta_{ j}\right)\right.\\
\times\left.\left[\Gamma_{3j }\left(\phi-\frac{\omega A_{\parallel}}{k_{\parallel}c}\right)+\Gamma_{4 j}\frac{k_{\mathrm B}T_{\perp j}}{q_{j}}\frac{\delta B_{\parallel}}{B_0}\right]\right\},\label{wperpmoment} 
\end{multline}
\begin{multline}
\delta p_{\parallel j}=q_{j}n_{0j}\frac{1}{R_j}\left\{\vphantom{\frac{A}{A}}\Gamma_{0j}\frac{\omega A_{\parallel}}{k_{\parallel}c}-\phi-\left[2R_{j}\zeta_{j}^2\left( 1+\zeta_{j}Z_{j}\right)+\Delta_{j}\right]\right.\\
\times\left.\left[\Gamma_{0j}\left(\phi-\frac{\omega A_{\parallel}}{k_{\parallel}c}\right)+\Gamma_{1j}\frac{k_{\mathrm B}T_{\perp j}}{q_{j}}\frac{\delta B_{\parallel}}{B_0}\right]\right\},\label{wparallelmoment}
\end{multline}
where we evaluate the integrals over $v_{\perp}$ in terms of the modified Bessel functions $I_m$ with $\Gamma_{0j}\equiv I_0(\lambda_j)e^{-\lambda_j}$, $\Gamma_{1j}\equiv \left[I_0(\lambda_j)-I_1(\lambda_j)\right]e^{-\lambda_j}$,  $\Gamma_{2j}\equiv 2\left[I_0(\lambda_j)-I_1(\lambda_j)\right]e^{-\lambda_j}$,  
$\Gamma_{3j}\equiv \Gamma_{0j}+\lambda_{j}\Gamma_{0j}^{\prime}$, and $\Gamma_{4j}\equiv 2\Gamma_{1j}+\lambda_{j}\Gamma_{1j}^{\prime}$
with the argument $\lambda_j\equiv k_{\perp}^2w_{\perp j}^2/2\Omega_{j}^2$.
We express the integrals over $v_{\parallel}$ in terms of the standard plasma dispersion function \citep{fried61}
\begin{equation}
Z_j\equiv \frac{1}{\sqrt{\pi}}\int _{\mathcal C} \frac{e^{-t^2}}{t-\zeta_j}\mathrm dt
\end{equation}
along the Landau contour $\mathcal C$, where $\zeta_j\equiv \omega/|k_{\parallel}|w_{\parallel j}$.

With these relations, the condition of quasi-neutrality, the parallel component of Amp\`ere's law, and the  perpendicular components of Amp\`ere's law ($\nabla_{\perp}\delta B_{\parallel}=(4\pi/c) \hat{\vec z}\times \delta \vec j$), where $\delta \vec j$ is the fluctuating part of the current density, yield:
\begin{align}
\sum\limits_j q_j\delta n_j &= 0.\label{qn} \\
k_{\perp}^2A_{\parallel}-\frac{4\pi}{c}\sum \limits_j q_j n_{0j} \delta U_{\parallel j} &=0,
\end{align}
and
\begin{multline}
\frac{\delta B_{\parallel}}{B_0}+\frac{4\pi}{B_0^2}\sum\limits_{j}q_jn_{0j}\left\{ \Gamma_{1j}\frac{\omega A_{\parallel}}{k_{\parallel}c}-\left(R_j\zeta_j Z_j+\Delta_j \right)\right. \\
\left.\times \left[\Gamma_{1j}\left(\phi-\frac{\omega A_{\parallel}}{k_{\parallel}c}\right)+\Gamma_{2j} \frac{k_{\mathrm B}T_{\perp j}}{q_j}\frac{\delta B_{\parallel}}{B_0}\right]\right\}=0,\label{perpamp} 
\end{multline}

As shown by \citet{howes06}, we can write Equations~(\ref{qn}) through (\ref{perpamp}) as
\begin{equation}\label{Dmat}
\begin{pmatrix}
A & A-B & C \\
A-B & A-B-\frac{P}{\bar{\omega}^2} & C+E \\
C & C+E & D-\frac{2}{\beta_{\perp \mathrm p}}
\end{pmatrix}
\begin{pmatrix}
\phi \\
-\frac{\omega A_{\parallel}}{k_{\parallel}c}\\
\frac{k_{\mathrm B}T_{\perp\mathrm p}}{q_{\mathrm p}}\frac{\delta B_{\parallel}}{B_0}
\end{pmatrix}=0
\end{equation}
 for $n_{0\mathrm p}=n_{0\mathrm e}$, where
\begin{align}
A&\equiv 1+\left(R_{\mathrm p}\zeta_{\mathrm p}Z_{\mathrm p}+\Delta_{\mathrm p}\right)\Gamma_{0\mathrm p}\\ \nonumber
&+\frac{T_{\perp\mathrm p}}{T_{\perp\mathrm e}}\left[1+\left(R_{\mathrm e}\zeta_{\mathrm e}Z_{\mathrm e}+\Delta_{\mathrm e}\right)\Gamma_{0\mathrm e}\right],\\
B&\equiv 1-\Gamma_{0\mathrm p}+\frac{T_{\perp\mathrm p}}{T_{\perp \mathrm e}}\left(1-\Gamma_{0\mathrm e}\right),\\
C&\equiv \left(R_{\mathrm p}\zeta_{\mathrm p}Z_{\mathrm p}+\Delta_{\mathrm p}\right)\Gamma_{1\mathrm p}-\left(R_{\mathrm e}\zeta_{\mathrm e}Z_{\mathrm e}+\Delta_{\mathrm e}\right)\Gamma_{1\mathrm e},\\
D&\equiv \left(R_{\mathrm p}\zeta_{\mathrm p}Z_{\mathrm p}+\Delta_{\mathrm p}\right)\Gamma_{2\mathrm p}+\frac{T_{\perp\mathrm e}}{T_{\perp\mathrm p}}\left(R_{\mathrm e}\zeta_{\mathrm e}Z_{\mathrm e}+\Delta_{\mathrm e}\right)\Gamma_{2\mathrm e},\\
E&\equiv \Gamma_{1\mathrm p}-\Gamma_{1\mathrm e},\\
P&\equiv \lambda_{\mathrm p}+\frac{1}{2}\beta_{\parallel\mathrm p}\Delta_{\mathrm p}\left(1-\Gamma_{0\mathrm p}\right)\nonumber\\
&+\frac{T_{\perp\mathrm p}}{T_{\perp\mathrm e}}\frac{1}{2}\beta_{\parallel\mathrm e}\frac{m_{\mathrm p}}{m_{\mathrm e}}\Delta_{\mathrm e}\left(1-\Gamma_{0\mathrm e}\right),
\end{align}
and $\bar{\omega}\equiv \omega/|k_{\parallel}|v_{\mathrm A}$.
Setting the determinant of the matrix in Equation~(\ref{Dmat}) to zero leads to the nontrivial solutions of the gyrokinetic dispersion relation, which fulfill
\begin{equation}\label{disp}
\left(\frac{PA}{\bar{\omega}^2}-AB+B^2\right)\left(\frac{2A}{\beta_{\perp\mathrm p}}-AD+C^2\right)=\left(AE+BC\right)^2.
\end{equation}
The term on the right side of Equation~(\ref{disp}) represents coupling terms that can be neglected for long wavelengths. In the long-wavelength limit  ($\lambda_{j}\ll 1$), we apply the approximations $\Gamma_{0j}\simeq 1-\lambda_j$, $\Gamma_{1j}\simeq 1-3\lambda_j/2$, and $\Gamma_{2j}\simeq 2-3\lambda_j$. Furthermore, we neglect all terms $\sim m_{\mathrm e}/m_{\mathrm p}$ given the assumed constrains on the temperature ratios. 

The first term on the left side of Equation~(\ref{disp}) represents the Alfv\'en solution, and the second term represents the slow-mode solution. 
In the long-wavelength limit, the Alfv\'en branch reduces to $P=\bar{\omega}^2\lambda_{\mathrm p}$, which leads to the Alfv\'en dispersion relation
\begin{equation}\label{alfven}
\omega=\pm k_{\parallel}v_{\mathrm A}\left(1+\frac{\beta_{\parallel\mathrm p}}{2}\Delta_{\mathrm p}+\frac{\beta_{\parallel e}}{2}\Delta_{\mathrm e}\right)^{1/2}.
\end{equation}
If the expression in parentheses in Equation~(\ref{alfven}) becomes negative, this root of the dispersion relation describes the firehose instability with the instability criterion 
\begin{equation}\label{firehose}
\beta_{\parallel\mathrm p}-\beta_{\perp\mathrm p}+\beta_{\parallel\mathrm e}-\beta_{\perp\mathrm e}>2.
\end{equation}
 
The slow-mode branch can be analytically simplified in two limits. 
In the low-$\beta_{\perp\mathrm p}$ limit, the slow-mode dispersion relation becomes $A=0$. Using the expansion
 \begin{equation}
Z_{\mathrm p}\simeq i\sqrt{\pi}e^{-\zeta_{\mathrm p}^2}-\frac{1}{\zeta_{\mathrm p}}-\frac{1}{2\zeta_{\mathrm p}^3}-\frac{3}{4\zeta_{\mathrm p}^5}
\end{equation}
and $\zeta_{\mathrm e}Z_{\mathrm e}\ll 1$ leads to
\begin{equation}\label{DRIA}
\frac{T_{\parallel\mathrm p}}{T_{\parallel\mathrm e}}+i\sqrt{\pi}\frac{\omega}{|k_{\parallel}|w_{\parallel\mathrm p}}e^{-\omega^2/k_{\parallel}^2w_{\parallel\mathrm p}^2}-\frac{k_{\parallel}^2w_{\parallel\mathrm p}^2}{2\omega^2}-\frac{3k_{\parallel}^4w_{\parallel \mathrm p}^4}{4\omega^4}=0.
\end{equation}
Under the assumption that $T_{\parallel\mathrm e}\gg T_{\parallel\mathrm p}$ and $\gamma\ll\omega_{\mathrm r}$, the real part of Equation~(\ref{DRIA}) leads to Equation~(\ref{omegaIA}), and the imaginary part of Equation~(\ref{DRIA}) leads to Equation~(\ref{gammaIA}).

In the high-$\beta_{\perp\mathrm p}$ limit, the slow-mode dispersion relation becomes $2/\beta_{\perp\mathrm p}=D$. Using the expansion $Z_{\mathrm p}\simeq i\sqrt{\pi}$ and $\zeta_{\mathrm e}\ll \zeta_{\mathrm p}$, this solution leads to the dispersion relation of the NP mode, $\omega_{\mathrm r}=0$ and Equation~(\ref{gammaNP}).

\section{Appendix~B\\Derivation of Moment Fluctuations in the Gyrokinetic Approximation}\label{app_PBS}

The perpendicular component of Amp\`ere's law in gyrokinetics implies a pressure balance in the form of
\begin{equation}\label{gyroPB}
\nabla_{\perp}\left(\frac{B_0\delta B_{\parallel}}{4\pi}+\delta \vec P_{\perp}\right)=0,
\end{equation}
where 
\begin{equation}
\delta \vec P_{\perp}\equiv \sum\limits_jm_j\int \left\langle \vec v_{\perp}\vec v_{\perp}h_j\right\rangle_{\vec r}\mathrm d^3v
\end{equation}
is the perpendicular pressure tensor and $\left\langle\cdot\right\rangle_{\vec r}$ is the ring average at fixed position $\vec r$.

For a given solution to the dispersion relation, Equation~(\ref{Dmat}) provides the polarization relations that connect the three amplitudes $\phi$, $A_{\parallel}$, and $\delta B_{\parallel}$:
\begin{equation}
\begin{pmatrix}
\phi\\
\frac{\omega A_{\parallel}}{k_{\parallel}c}
\end{pmatrix}=
\begin{pmatrix}
L_{\phi}\\
L_A
\end{pmatrix}\frac{k_{\mathrm B}T_{\perp\mathrm p}}{q_{\mathrm p}}\frac{\delta B_{\parallel}}{B_0}, 
\end{equation}
where
\begin{equation}\label{Lphi}
L_{\phi}= \frac{\left(D-\frac{2}{\beta_{\perp\mathrm p}}\right)\left(A-B-\frac{P}{\bar{\omega}^2}\right)-(C+E)^2}{\frac{PC}{\bar{\omega}^2}+AE-BE}
\end{equation}
and
\begin{equation}\label{LA}
L_A= \frac{\left(D-\frac{2}{\beta_{\perp\mathrm p}}\right)\left(A-B\right)-C(C+E)}{\frac{PC}{\bar{\omega}^2}+AE-BE}.
\end{equation}

With Equations~(\ref{nmoment}) through (\ref{wparallelmoment}),  (\ref{Lphi}), and (\ref{LA}), we find for the normalized fluctuations in the three lowest velocity moments:
\begin{multline}\label{xiGK}
\xi_j=\tau_j\left\{\vphantom{\frac{A}{A}}\Gamma_{0j}L_A-L_{\phi}-\left(R_j\zeta_jZ_j+\Delta_j\right)\right.\\
\left.\times\left[\Gamma_{0j}\left(L_{\phi}-L_A\right)+\frac{\Gamma_{1j}}{\tau_j}\right]\right\},
\end{multline}
\begin{multline}\label{chiGK}
\chi_j=\tau_j\left\{\frac{1}{2}\frac{k_{\parallel}v_{\mathrm A}}{\omega} \frac{m_{\mathrm p}}{m_j}\beta_{\parallel j}\Delta_j\left(\Gamma_{0j}-1\right)L_A\right .\\
\left. -\frac{\omega}{k_{\parallel} v_{\mathrm A}}R_j\left(1+\zeta_jZ_j\right) \left[\Gamma_{0j}\left(L_{\phi}-L_A\right)+\frac{\Gamma_{j1}}{\tau_j}\right]\right\},
\end{multline}
\begin{multline}
\alpha_{\perp j}= \tau_j\beta_{\perp j}\left\{\vphantom{\frac{A}{A}}\Gamma_{3j}L_A-L_{\phi}-\left(R_j\zeta_jZ_j+\Delta_j\right)\right.\\
\left.\times\left[\Gamma_{3j}\left(L_{\phi}-L_A\right)+\frac{\Gamma_{4j}}{\tau_j}\right]\right\},
\end{multline}
\begin{multline}\label{alphaparGK}
\alpha_{\parallel j}= \tau_j \beta_{\parallel j}\left\{\vphantom{\frac{A}{A}}\Gamma_{0j}L_A-L_{\phi}\right.\\
\left.-\left[2R_j\zeta_j^2\left(1+\zeta_jZ_j\right)+\Delta_j\right]\right. \\
\left.\times \left[\Gamma_{0j}\left(L_{\phi}-L_A\right)+\frac{\Gamma_{1j}}{\tau_j}\right]\right\},
\end{multline}
where
\begin{equation}
\tau_j\equiv \frac{q_jT_{\perp \mathrm p}}{q_{\mathrm p}T_{\perp j}}.
\end{equation}
We then calculate the factor $\psi$ with Equation~(\ref{psieq}). We evaluate Equations~(\ref{xiGK}) through (\ref{alphaparGK}) after solving Equation~(\ref{Dmat}) numerically (i.e., without approximating $\Gamma_{ij}$ and $Z_j$) for $\omega$, $L_{\phi}$, and $L_A$.
In the marginally stable case for the mirror-mode instability, the NP mode fulfills $\omega=0$, $L_{\phi}=-C/A$,  $L_A=0$, and Equation~(\ref{mirror}), after replacing the inequality sign in Equation~(\ref{mirror}) with an equal sign.

\section{Appendix C\\MHD Polarization Relations}\label{app_MHD}

Linearizing the ideal adiabatic MHD equations for a proton fluid,
\begin{equation}
\diffp{n_{\mathrm p}}{t}+\nabla\cdot \left(n_{\mathrm p}\vec U\right) =0, \label{conti} 
\end{equation}
\begin{equation}
\diffp{\vec U}{t}+\left(\vec U\cdot \nabla\right)\vec U=-\frac{\nabla p}{n_{\mathrm p}m_{\mathrm p}}+\frac{1}{4\pi n_{\mathrm p}m_{\mathrm p}}\left[\left(\nabla\times \vec B\right)\times \vec B\right],
\end{equation}
\begin{equation}
\diffp{\vec B}{t}=\nabla\times\left(\vec U\times \vec B\right),
\end{equation}
\begin{equation}
pn_{\mathrm p}^{-\kappa}=\mathrm{constant},\label{adia}
\end{equation}
where $\vec U$ is the fluid velocity and $p$ is the pressure, leads to slow-magnetosonic wave solutions with the dispersion relation given by Equations~(\ref{omegaMHD}) and (\ref{Cminus}). Combining Equations~(\ref{conti}) through (\ref{adia}) with Equations~(\ref{xi}), (\ref{chi}), and (\ref{psi}) leads to \citep{verscharen16}
\begin{equation}\label{xiMHD}
\xi_{\mathrm{MHD}}=\frac{C_-^2}{C_-^2-\frac{\kappa}{2}\beta_{\parallel \mathrm p}\cos^2\theta},
\end{equation}
\begin{equation}
\chi_{\mathrm{MHD}}=\frac{C_-\cos\theta}{C_-^2-\frac{\kappa}{2}\beta_{\parallel \mathrm p}\cos^2\theta}\frac{\kappa}{2}\beta_{\parallel\mathrm p},
\end{equation}
and
\begin{equation}
\psi_{\mathrm{MHD}}=\kappa \beta_{\parallel\mathrm p}\xi_{\mathrm{MHD}}.
\end{equation}
We note that $\psi_{\mathrm{MHD}}$ describes the effect of the total isotropic pressure instead of the proton partial pressure alone.

\bibliographystyle{aasjournal}
\bibliography{slow_modes}

\end{document}